\documentclass[twocolumn,aps,prl,groupedaddress,amsmath,amssymb,longbibliography]{revtex4-1}
\usepackage{url}
\usepackage{ulem}
\usepackage[colorlinks=true, linkcolor=blue,urlcolor=blue,anchorcolor=blue,citecolor=blue,bookmarksnumbered]{hyperref}
\usepackage{graphicx}
\usepackage{dcolumn}
\usepackage{bm}
\usepackage{verbatim}

\begin{document}
	
	\title{Topological Quantum Transducers in a Hybrid Rydberg Atom System}
	\author{Pei-Yao~Song$^{1}$}\email[]{These authors contributed equally to this work.}\author{Jin-Lei~Wu$^{1}$}\email[]{These authors contributed equally to this work.} \author{Weibin~Li$^{2}$}\email[]{weibin.li@nottingham.ac.uk} \author{Shi-Lei~Su$^{1,3}$}\email[]{slsu@zzu.edu.cn}
	\affiliation{$^{1}$School of Physics, Zhengzhou University, Zhengzhou 450001, China\\
		$^2$School of Physics and Astronomy, and Centre for the Mathematics and Theoretical Physics of Quantum Non-equilibrium Systems, The University of Nottingham, Nottingham NG7 2RD, United Kingdom\\
		$^3$Institute of Quantum Materials and Physics, Henan Academy of Science, Henan 450046, China}
	
	\begin{abstract}
		 We propose a topological transport platform for microwave-to-optical conversion at the single-photon level in a Rydberg atom-cavity setting. This setting leverages a  hybrid dual-mode Jaynes-Cummings (JC) configuration, where a microwave resonator couples an optical cavity mediated by a Rydberg atom ensemble. Our scheme uniquely enables the formation of Fock-state lattices (FSLs), where photon hopping rates depend on photon numbers in individual sites. We identify an inherent zero-energy mode corresponding to the dark state of the dual-mode JC model. This enables to build a high-efficiency single-photon transducer, which realizes topologically protected photon transport between the microwave and optical modes. Crucially, we show analytically that the FSL features continuous variations of the winding number. Our work establishes a robust mechanism for efficient quantum transduction in synthetic dimensions  and opens avenues for exploring topological physics with continuous winding numbers in the atom-cavity system. 
	\end{abstract}
	\maketitle
\textbf{\textit{Introduction.}}---Quantum transduction between microwave (MW) and optical domains at the single-photon level is a critical requirement for building scalable quantum networks~\cite{Xie2025NP}. The quantum transducer bridges local quantum processors with fiber-optic channels for remote communication~\cite{Tu2022,Kumar2023,Borowka2024,Meesala2024NP,xuHan2021Optica,wjWu2025LPR,Petrosyan2019NJP}. In the past two decades, significant developments have been achieved in optomechanical and electro-optic systems~\cite{Cernotik2018PRL,Rueda2019npjQuanInf}. The remaining challenge is to  overcome  low conversion efficiencies due to weak nonlinear effects and decoherent thermal noise~\cite{Forsch2020,Jiang2020NC,Rochman2023,Chen2023NC,vanThiel2025}.  Topologically protected transport is a promising approach as it is robust against disorder and thus potentially improves conversion efficiency ~\cite{Kraus2012,FMei2018PRA,Jurgensen2021Nature,Cheng2022NC,Citro2023NRP,Song2024SciAdv,Wu2024NC,JNZhang2024PRB}. Existing schemes are either suffered from disorders of the multi-site systems~\cite{Leseleuc2019Science,Kiczynski2022Nature,Youssefi2022Nature,chzhao2024PRB,Ghosh2025PRResearch}, or operated in the classical regime~\cite{jsHan2018PRL}, and thus possess  limitations in realizing quantum transduction.

	\begin{figure}[htp!]\centering
		\includegraphics[width=\linewidth]{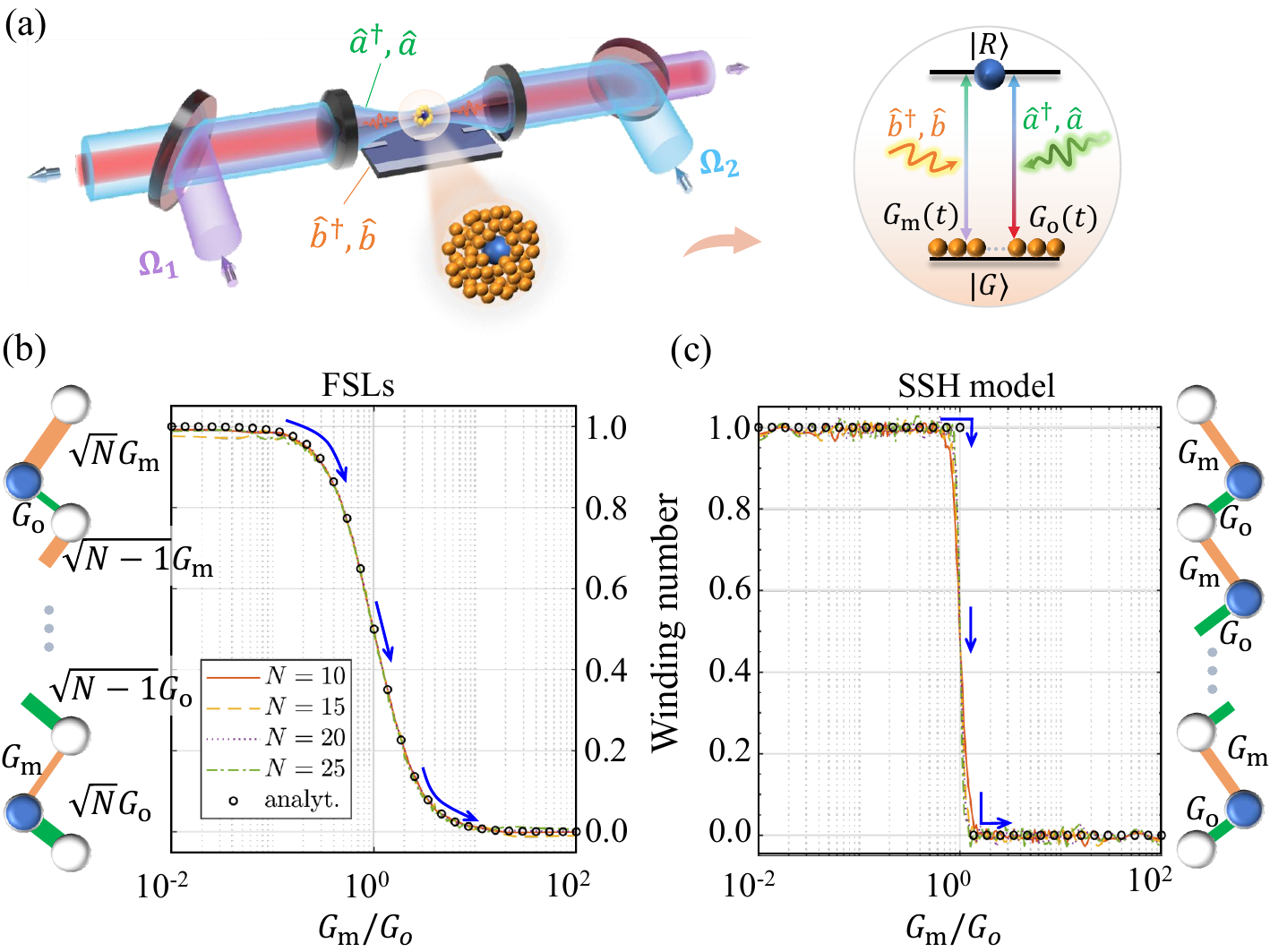}
		\caption{\textbf{Topological quantum transducer with continuous winding number.} (a)~Scheme of Rydberg-superatom quantum transducer based on 4WM, where atoms in the superatom are coupled simultaneously to an optical cavity (${\hat a}^\dagger$, ${\hat a}$) and a MW resonator (${\hat b}^\dagger$, ${\hat b}$), assisted by two counterpropagating 297~nm- (Rabi frequency $\Omega_1$) and 481~nm-wavelength (Rabi frequency $\Omega_2$) lasers; $G_m$~($G_o$) is the coupling strength between the superatom spin and the MW (optical) mode.  Topological pumping based on (b) FSL~and (c) SSH model. The winding number changes continuously and discretely in (b) and (c), respectively. This is implemented when the ratio $G_m/G_0$ is modulated continuously from 0 to $\infty$. Winding number data with different excitation numbers $N$ are obtained by numerically measuring the time-averaged chiral displacements. See text for details.}\label{fig1}
	\end{figure} 
	On the other hand, Fock-state lattices~(FSLs) are an ideal setting for studying topological phases of light in the quantum realm~\cite{DWWang2016PRL,HCai2020NSR,JLYuan2021APLphotonics}. FSLs consist of collective atom-photon states with finite photon-number. Such synthetic lattice is flexible in geometry, while hopping along the synthetic lattice depends on the number of photons. Recent experiments have demonstrated exotic topological properties of quantized light with superconducting circuits~\cite{JFDeng2022Science,Ehrhardt2024science,JJZhang2025}. Despite promising for quantum information applications~\cite{DWWang2016PRL, Saugmann2023PRA,JQLiao2025PRResearch}, quantum interfaces based on FSLs are largely unexplored, mainly due to two challenges: constructing FSLs incorporating light modes across distinct bands, and indentifying topological invariants in translation-symmetry breaking FSLs.

	In this work, we propose a light-matter hybrid FSL scheme for efficient microwave-to-optical (MTO) single-photon transducers. This is realized in a platform where a Rydberg atom ensemble is coupled simultaneously to a microwave resonator and an optical cavity [see Fig.~\ref{fig1}(a)]. This creates a dual-mode FSL where the underlying frequencies differ by several orders of magnitude. Temporal modulation of the light-atom coupling drives disorder-resilient photon transport protected by a topological zero-energy mode of the FSL. Uniquely, we find that the winding number of the system evolves continuously with the modulation, rather than jumps between discrete values in SSH models~\cite{Leseleuc2019Science} [Figs.~\ref{fig1}(b) and ~\ref{fig1}(c)]. We identify that highly efficient conversion takes place with a wide range of initial states including Fock states, coherent states, and squeezed vacuum states. Our scheme overcomes the challenges in implementing topological quantum transducers, and provides a scheme for probing continuous winding number dynamics in the hybrid FSLs. It furthermore opens quantum technological applications that enable hardware integration in quantum network with topological protection~\cite{Krastanov2021PRL,Meesala2024PRX}.

		\textbf{\textit{Rydberg-superatom quantum transducer.}}---We start by presenting our hybrid quantum system schematically depicted in Fig.~\ref{fig1}(a). An ensemble with $N_a$ cold atoms is confined in an optical microtrap. They couple simultaneously to a superconducting coplanar waveguide resonator\cite{brown2025arxivHogan} and an optical cavity through a laser-assisted diamond level configuration, based on recent experimental implementation~\cite{Kumar2023}; See {\bf Supplementary Material~(SM)}~\cite{SM} for details. Through four-wave mixing (4WM), the atom-light coupling yields an effective dual-mode Jaynes-Cummings~(JC) Hamiltonian~($\hbar=1$) $\hat{H}_{\rm eff}=G_m/\sqrt{N_a}|r_2\rangle \langle g|\hat{b}+G_o/\sqrt{N_a}|r_2\rangle \langle g|\hat{a}+\rm{H.c.}$, where $|g\rangle$=$|5S_{1/2}\rangle$ is an atomic hyperfine ground state while $|r_2\rangle$=$|71S_{1/2}\rangle$ is a high-lying Rydberg state. $G_m$ and $G_o$ are collective coupling strengths of the atomic ensemble to the MW mode~(creation and annihilation operators $\hat{b}^{\dagger}$ and $\hat{b}$) and optical mode~(operators $\hat{a}^{\dagger}$ and $\hat{a}$), respectively. We will work in the blockade regime, i.e. ensemble size is smaller than the Rydberg blockade radius~\cite{Lukin2021PRL,Yang2022Optica,XQShao2024APR} or in the adiabatic coupling~\cite{SM}. A Rydberg superatom is formed with the excitation number $\leq1$. In this situation, we obtain a superatom-enhanced dual-mode JC model~[Fig.~\ref{fig1}(a)]
	\begin{equation}\label{eq1}
	\hat{H}_{D}=G_{\rm m}|R\rangle\langle G|\hat{b}+G_{\rm o}|R\rangle\langle G|\hat{a}+\rm{H.c}.,
	\end{equation} 
	with the collective ground state $|G\rangle=|g_1g_2g_3\cdots g_{N_a}\rangle$ and Rydberg state $|R\rangle=1/\sqrt{N_a}\sum_j^{N_a}|g_1g_2g_3\cdots r_{2,j} \cdots g_{N_a}\rangle$. This dual-mode JC model can be solved analytically, providing insights into   the dynamic evolution of the quantum system~\cite{SM}.
	
We consider an initial condition, where the MW mode, superatom and the optical mode are prepared in $N$ photon state (Fock state $|N_m\rangle$), $|G\rangle$ and vacuum state $|0_o\rangle$, respectively. The excitation-number conservation of the dual-mode JC model constraints the system in a chainwise FSL holding $2N+1$ sites represented by joint MW-superatom-photon product states. The Hamiltonian turns out to be
	\begin{equation}\label{eq2}
	\hat{H}^{(N)}_{\rm FSL}=\sum_{j=1}^{N}u_j|2j-1\rangle\langle2j|+v_j|2j\rangle\langle2j+1|+\rm{H.c.},
	\end{equation}
	with $|2j-1\rangle=|(N-j+1)_m,G,(j-1)_o\rangle$, $|2j\rangle=| (N-j)_m,R,(j-1)_o\rangle$, and $|2j+1\rangle=|(N-j)_m,G,j_o\rangle$. $u_j=G_{m}\sqrt{N-j+1}$ and $v_j=G_o\sqrt{j}$ are photon-number--dependent hopping rates between two nearest-neighbor sites. Importantly, Hamiltonian~(\ref{eq2}) supports a zero-energy mode
	$|\phi_0\rangle \sim\sum_{j=0}^N\sqrt{\frac{N!}{(N-j)!j!}}(G_{\rm o})^{N-j}(-G_{\rm m})^j|2j+1\rangle$, which connects two edge sites of the FSL, $|1\rangle=|N_m,G,0_o\rangle$ and $|2N+1\rangle=|0_m,G,N_o\rangle$. This permits topological pumping of photons from the MW mode to the optical mode. This is the key to building our quantum transducers.
	
	\textbf{\textit{Continuous winding number.}}---In existing FSL-based protocols~\cite{HCai2020NSR,JLYuan2021APLphotonics,JFDeng2022Science,CHWu2023PRA,TTian2024PRB,ZGuan2024PRApplied,MPeng2025PRL,Yang2024NC}, the chainwise FSLs~[zig-zag chain in Fig.~\ref{fig1}(b)] are typically described by the SSH model~[Fig.~\ref{fig1}(c)]. In the FSL model, however,  the coupling  is related to the number of photons. It breaks the translational invariance, which  different from the schemes based on the SSH model. The effect due to the broken translational symmetry on the topological number has not been discussed in this regime. In Fig.~\ref{fig1}(b), the topological properties of the system without the translation invariance and the range of coupling ratios corresponding to trivial and non-trivial winding numbers are shown. In the SSH model, quantum states are transferred via a zero-energy mode that becomes adiabatically delocalized across the chain. In contrast, the FSLs here host a domain wall (a localized defect state between two distinct topological phases) that can be swept across the lattice to transfer a quantum state. The translation-symmetry breaking makes it difficult to determine the topological invariants of FSLs, which relies on the Berry curvature of the momentum-space bands. 
    
  To overcome this limitation, we introduce a dynamical, mean chiral displacement~(MCD) method~\cite{Cardano2017NC}, which is capable of determining winding numbers of chiral-symmetry lattices~\cite{WCai2019PRL,Aravena2024PRL}.  Similar to the standard SSH model, the FSLs in our setting hold the chiral symmetry, $\Gamma_c\hat{H}^{(N)}_{\rm FSL}\Gamma_c=-\hat{H}^{(N)}_{\rm FSL}$ with $\Gamma_c\equiv\sum_{j=1}^{N}|2j-1\rangle\langle2j-1|-|2j\rangle\langle2j|$.
	For comparison, we consider a SSH chain containing $2N+1$ sites [Fig.~\ref{fig1}(c)] with intra- and inter-cell hopping rates $G_m$ and $G_o$, respectively, described by Hamiltonian Eq.~\eqref{eq2} but with modified $u_j=G_m$ and $v_j=G_o$. Here the chiral displacement is defined as $P_d(t)=\sum_{j=1}^Nj[P_{2j-1}(t)-P_{2j}(t)]$, where $P_{j'}(t)$ denotes the instantaneous
	excitation population of the $j'$-th site. The winding number is extracted through dynamic evolution, ${\mathcal W}=\lim_{\tau\rightarrow\infty}(2/\tau)\int_0^\tau P_d(t){\rm d}t$, where the system locates at an even-numbered site initially. 
	
			\begin{figure*}[t]\centering
		\includegraphics[width=\linewidth]{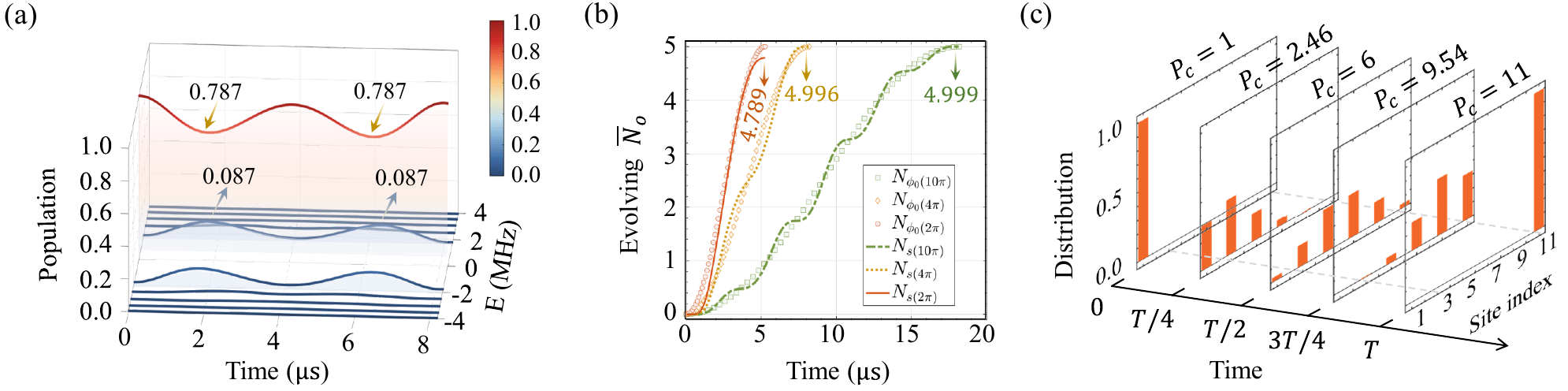}
		\caption{\textbf{Topological pumping for MTO photon conversion.} (a)~Energy spectrum versus time and population evolution of eigenstates when considering the initial state $|5_m,G,0_o\rangle$ and $T=8.2~{\rm \mu s}$. (b) Temporary evolution of ${\overline N}_o(t)$  with $T=5.26~{\rm \mu s}$, $8.2~{\rm \mu s}$, and $18.18~{\rm \mu s}$, corresponding to effective pulse area $A=2\pi$, $4\pi$, and $10\pi$, respectively. Photon number $N_{\phi_0}(A)$ and $N_{s}(A)$ are calculated based on the zero-mode state $|\phi_0(t)\rangle$ and density operator of the syste,, respectively. (c)~Population distribution on the eleven sites with $T=8.2~{\rm \mu s}$.}\label{fig2}
	\end{figure*}
Different from the SSH-based topological pumping protocols~\cite{FMei2018PRA,TTian2022PRL,WLiu2022PRA,Song2024SciAdv,Wu2024NC,JNZhang2024PRB,JKGuo2024PRA,JXHan2024PRApplied}, excitation is transported along the zero-energy mode from one edge of the chain to the other, for FSLs involving photon-number--dependent hopping rates. We find that the winding numbers by the MCD vary continuously with ratio $G_m/G_o$, as shown in Fig.~\ref{fig1}(b). Contrastively, for the SSH model the modulation of intersite hoppings from $G_m<G_o$ to $G_m>G_o$ indicates a phase transition from topological nontriviality to triviality, where the winding number changes from 1 to 0. These two winder numbers can be analytically calculated based on momentum-space bands or be numerically extracted by the MCD method in Fig.~\ref{fig1}(c). For general cases,  we obtain the winding number of FSLs analytically,
 \begin{equation}\label{eq3}
{\mathcal W}=\cos^2\left[\arctan\left(\frac{G_m}{G_o}\right)\right],
 \end{equation}
which is essentially different from the discrete values of the SSH model. This corresponds to a progressive displacement of the photon distribution center along the zero-energy mode when changing $G_m/G_o$, populating only on the odd-numbered sites. Uniquely, we characterize the position of photon distribution center in the zero-energy mode by
\begin{equation}\label{eq4}
P_c=2N(1-{\mathcal W})+1,
\end{equation}
 which is applicable in both the FSLs and the standard SSH model.

\textbf{\textit{MTO photon conversion.}}---To achieve the topological quantum transducer, we employ temporary modulation coupling fields, $G_m(t) = g \sin(\pi t / 2T)$ and $G_o(t) = g \cos(\pi t / 2T)$, where $T$ is the modulation duration. We take into account of experimentally feasible parameters in the 4WM fields~(see {\bf SM}~\cite{SM} for details). With these considerations, we set $g/2\pi=0.282$~MHz, which is the maximum coupling  between the superatom and the MW/optical mode. These choices enable the time-independent spectrum with eigenenergies $E_0=0$ and $E_{\pm j}=\pm\sqrt jg$, corresponding to time-dependent eigenstates $|\phi_0(t)\rangle$ and $|\phi_{\pm j}(t)\rangle$~(see {\bf SM}~\cite{SM} for analytical derivations), respectively. As shown in Fig.~\ref{fig2}(a), the constant eigenenergies avoids gap shrinking at $G_m(t)=G_o(t)$, facilitating faster topological pumping over the standard SSH lattice~\cite{FMei2018PRA,TTian2022PRL,WLiu2022PRA,Song2024SciAdv,Wu2024NC,JNZhang2024PRB,JKGuo2024PRA,JXHan2024PRApplied}.

The adiabatic evolution dominated by path along state $|\phi_0(t)\rangle$ is guaranteed by increasing the pumping duration. The duration threshold for a given $N$ can be moderately shortened,~e.g., $T_m=8.2~{\rm\mu s}$ for $N=5$. This is achieved by a small admixture of nonadiabatic cyclic transitions from state $|\phi_0(t)\rangle$ to $|\phi_{\pm j}(t)\rangle$, but finally returning to state $|\phi_0(t)\rangle$~[population oscillations in Fig.~\ref{fig2}(a)]~\cite{junjing2022pra,TTian2024PRB,JNZhang2024PRB}. In this situation, the effective pulse area $A$ is required to be an integer multiple of $2\pi$~({\bf SM}~\cite{SM}). Optimally one would increase the pulse area because a smaller $A$ induces excitation of states $|\phi_{\pm j}(t)\rangle$ with larger-$j$ in the dynamics, which hinders $|\phi_0(t)\rangle$ returning to to itself~[Fig.~\ref{fig2}(b)].  The instantaneous average optical photon number is defined by ${\overline N}_o(t)={\rm tr}[{\hat \rho}(t){\hat a}^\dagger{\hat a}]$, ${\hat \rho}(t)$ being the instantaneous density operator of the hybrid system. The density operator ${\hat \rho}(t)$ is obtained by numerically the solving Lindblad master equation ({\bf SM~\cite{SM}}). The temporary modulation changes the ratio $G_m/G_o $ from $0\rightarrow\infty$, which continuously changes the winding number  from 1 to 0. This  progressively shifts the photon distribution center,  $P_c(t=0)=1\rightarrow P_c(t=T)=2N+1$, as shown in Fig.~\ref{fig2}(c). Such changes realizes the photon operation in the quantum transducer. 

\begin{figure}
\centering
	\includegraphics[width=\linewidth]{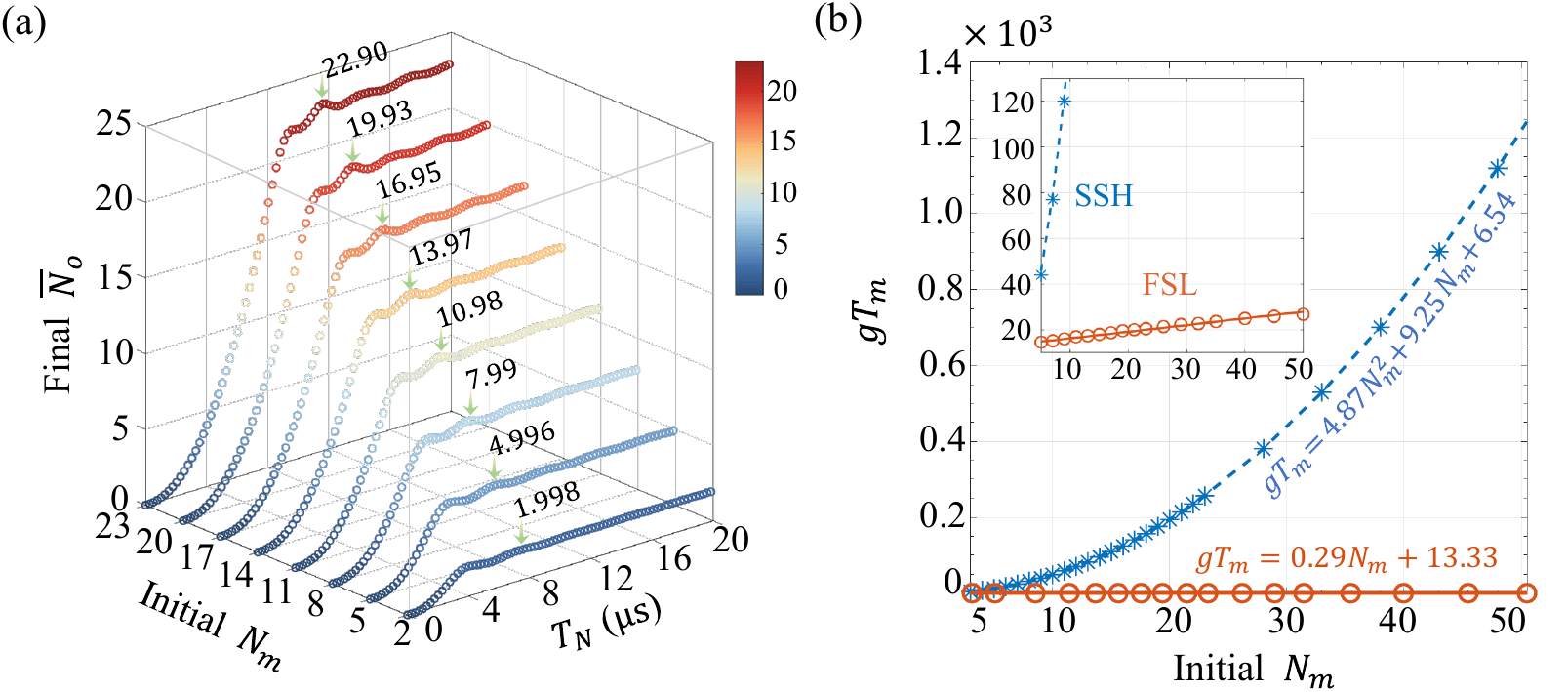}
	\caption{~(a) Final average optical photon number ${\overline N}_o(T)$ versus topological pumping duration $T_N$ with different $N_m$. The green arrows indicate the critical transfer time $T_m$  for different initial $N_m$. They take place at the first peak of the transmission fidelity. The number shows the final photon number.  For $N_m=5$, we obtain $T_m=8.2 ~\mu s$, which has been used as the example discussed  in the main text.  (b) Dimensionless time $gT_m$ versus initial $N_m$.  Circles (numerical): FSL model. Snowflake (numerical): SSH model at which the fidelity for different initial $N_m$ in the  reaches 99\%. Solid line (fitting): FSL model with $gT_m=0.29N_m+13.33$. Dashed line (fitting): SSH model with $gT_m=4.87N_m^2+9.25N_m+6.54$.} \label{fig3} 
\end{figure}
\textbf{\textit{Scalability and robustness.}}--- The duration threshold $T_m$ to finally obtain a sufficient optical photon number ${\overline N}_o$ increases linearly with the initial excitation number $N_m$ due to the extension of the FSL chain size [see Fig.~\ref{fig3}]. However, such an increase with $N$ is slow because the spectrum gap is constant $g$, independent of the lattice size, which is essentially different from the SSH model. For example, at $T_m=8.2~{\rm \mu s}$, highly efficient MTO photon conversion can be obtained even when increasing the excitation number to a certain degree~[Fig.~\ref{fig3}(a)].  We numerically obtain the critical time $T_m$ of different $N_m$ when the underlying transfer fidelity is above 99\%. It is found that $gT_m$ depends quadratically on $N_m$ for the SSH model. A unique  advantage of the FSL is that the energy gap remains constant throughout the transmission, whereas in the SSH model the gap narrows as the initial $N_m$ increases (see {\bf SM}~\cite{SM}).

 \begin{figure}\centering
	\includegraphics[width=\linewidth]{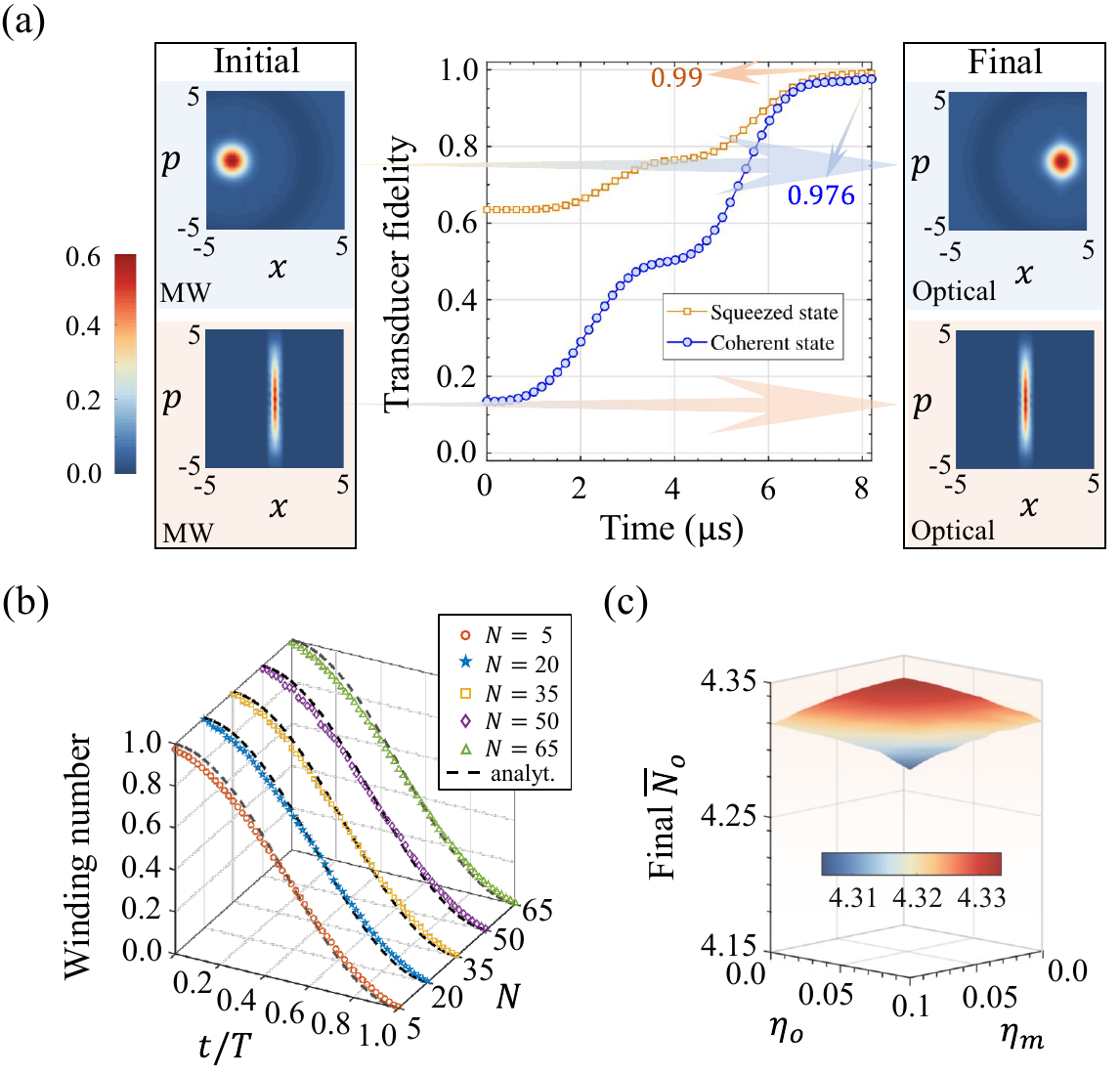}
	\caption{\textbf{Scalability and robustness of the topological quantum transducers.}   (a) Transducer fidelity evolution when the MW mode is initially in a coherent state (top, with $\alpha=1$) and a squeezed vacuum state (bottom, with $r=0.7$ and $\theta=0$). The Wigner functions of the initial MW photon state and the final optical photon state are shown on the leftmost and rightmost sides, respectively. The fidelity is defined by ${\mathcal F}(t)={\rm Tr}[{\hat\rho}(t)|\phi_{id}\rangle \langle\phi_{id}|]$, where the ideal state is $|\phi_{id}\rangle=|0_m,G,\alpha_o(\xi_o)\rangle$ with $|\alpha_o\rangle={\rm exp} (-\frac{1}{2}|\alpha|^2)\sum_{n=0}^{\infty}(-1)^n\frac{\alpha^n}{\sqrt{n!}}|n\rangle$ and $|\xi_o\rangle={\rm exp}[\frac{1}{2}(\xi^*{\hat a}^2-\xi{{\hat a}^{\dagger^2}})]|0_o\rangle$. (b)~Evolution of the winding number of the FSL with $\eta_m=\eta_o=0.1$. For  different excitation numbers,  the winding number changes gradually with time. (c)~Robust transducer against the disorder. We show the final ${\overline N}_o$ with initial $N_m=5$ and $T=8.2~{\rm \mu s}$. $\eta_m$ ($\eta_o$) is the disorder magnitude fluctuating coupling strength between the superatom and the MW (optical) mode: $G_{m(o)}\rightarrow G_{m(o)}[1+\epsilon_{m(o)}]$ with random samples $\epsilon_{m(o)}\in[-\eta_{m(o)},\eta_{m(o)}]$. Each data point on the surface plots is obtained through an ensemble average of 1001 sampings. In all panels, experimentally feasible decay rates of the superatom, MW mode, and optical mode are considered as $\Gamma_0/2\pi=3.6$~kHz~\cite{Meesala2024PRX,Sahu2023Science}, $\kappa_m/2\pi=2$~kHz~\cite{Pirkkalainen2015NC}, and $\kappa_o/2\pi=3.4$~kHz~\cite{Kongkhambut2022Science}, respectively.}\label{fig4}
\end{figure}
Benefited from this property, the present topological quantum transducer scales linearly with respect to the excitation number, enabling MTO photon conversions beyond Fock states. To illustrate this, we consider the MW mode initially in a coherent state $|\alpha_m\rangle={\rm exp} (-\frac{1}{2}|\alpha|^2)\sum_{n=0}^{\infty}\frac{\alpha^n}{\sqrt{n!}}|n\rangle$ ($\alpha$ is complex and $|\alpha|^2$ describes the average photon number) and a squeezed vacuum state $|\xi_m\rangle={\rm exp}[\frac{1}{2}(\xi^*{\hat b}^2-\xi{{\hat b}^{\dagger 2}})]|0_m\rangle$ ($\xi=re^{i\theta}$ is squeezing parameter with $r$ being the squeezing magnitude and $\theta$ determining the squeezing direction), respectively. By comparing Wigner functions [Fig.~\ref{fig4}(a)] of the initial and final states, the photons are faithfully converted from the MW to the photon mode. Fidelities of transferring the two photon states are shown in Fig.~\ref{fig4}(a). Both cases gain transducer fidelity $>0.91$ in the end of the operation. The infidelity mainly attributes to leakage to the nonadiabatic states.

Moreover the present quantum transducer is robust against disorder of the intersite hopping rates. This roots from the gap protection during topological pumping, manifesting as the robustness of the winding number [see Fig.~\ref{fig4}(b)]. We numerically evaluate winding numbers of the FSLs during topological pumping. To demonstrate the robustness, we add disorders to both modes via $G_{m(o)}[1+\epsilon_{m(o)}]$ with strength randomly sampled from $\epsilon_{m(o)}\in[-\eta_{m(o)},\eta_{m(o)}]$. Different excitation numbers and relatively larger disorder strength $\eta_m=\eta_o=0.1$ are then investigated. The results in Fig.~\ref{fig4}(b) show a high degree of coincidence between measured winding numbers and the analytic counterpart predicted by Eq.~\eqref{eq3}. The agreement is good even with strong disorder and for a large range of excitation numbers. Finally, our numerical simulations show that our scheme is robust against disordered parameters in Fig.~\ref{fig4}(c). The data shows that it is hardly affected by the fluctuations of the coupling strengths.

\textbf{\textit{Conclusion.}}---By leveraging a Rydberg-superatom--coupled hybrid MW-optics system, we have engineered synthetic FSLs with photon-number--dependent hoppings, enabling topological pumping through a zero-mode channel.  This gives a topological quantum transducer platform enabling efficient MTO photon conversion at the single-photon level. Crucially, this platform exhibits a continuous evolution of winding number, which is fundamentally distinct from conventional discrete phase transitions during topological pumping, resulting in progressive displacement of the photon distribution center via (quasi)-adiabatic modulation. We have shown that  high-fidelity transduction ($>91\%$) for various input states (Fock, coherent, and squeezed vacuum states). Robustness against the disorder (up to $10\%$ coupling-strength fluctuations) and scalability to multi-photon regimes stem from the protected transport by excitation-number--independent energy gap. This work establishes a topologically protected quantum interface that connects superconducting microwave processors with optical communication channels~\cite{Hogan2020PRL,Petrosyan2024EPJQT}. The setting could be integrated with other topological photonic, optomechanical, or circuit-based devices, thereby providing a promising platform for exploring exotic topological phenomena and realizing robust hybrid quantum networks.


\textbf{\textit{Acknowledgments.}}---This work was supported by the National Natural Science Foundation of China (Grants No. 62571494, No. 12274376, No. 12575032, No. 12304407), and the China Postdoctoral Science Foundation (No. 2024M762973 and No. 2023TQ0310), and a major science and technology project of Henan Province under Grant No. 221100210400, and the Natural Science Foundation of Henan Province under Grant No. 232300421075.

	\bibliography{myref}~\\
	
\end{document}